\documentclass[twocolumn, showpacs, preprintnumbers,amsmath,amssymb, nofootinbib]{revtex4-1}
\usepackage[dvips]{graphicx}
\usepackage{graphics}

\usepackage{color}

\renewcommand{\vec}[1]{\overrightarrow{#1}}

\begin{document}

\title{
Ambiguity in deriving Larmor's formula from the LAD equation of motion
}

\author{Sofiane Faci$^*$  and Jos\'e A. Helay$\ddot e$l-Neto}
\email{sofiane@cbpf.br} \email{helayel@cbpf.br}
\affiliation{
Centro Brasileiro de Pesquisas F\'isicas, CBPF,
Rua Xavier Sigaud, 150, 
Rio de Janeiro - Brazil
}

\date{\today}

\begin{abstract}
The power radiated by a moving charge is given by Larmor's formula which can be derived by integrating the Li\'enard-Wiechert  potential over the whole past history of the charge. However, extracting the same result from the Lorent-Abraham-Dirac (LAD) equation of motion is problematic. This is well known for uniform proper acceleration for which case the radiation force vanishes and hence the very origin of the radiated energy is not clear, leading to an energy balance paradox. 
The purpose of this letter is first to evince that this problem occurs in the general case and not only for uniform acceleration. Second we show that the widely accepted treatment based on the bound field technique cannot fix the energy balance discrepancy. Indeed the related Schott term is not a legitimate four-momentum for being indefinite and not conserved.
\end{abstract}

\maketitle


\paragraph*{Introduction.}

The Lorentz-Abraham-Dirac (LAD) equation is known to suffer from two pathologies \cite{Rohrlick-book, jackson1999classical,  
Hammond-2010}. 
First, self-acceleration (or \textit{runaway}) is due to unstable and divergent solutions, a free charge can spontaneously start accelerating and emitting radiation in an exponential rate.  Second, there is the pre-acceleration behaviour for the acceleration always precedes the external force acting on the charge, leading to causality violation. These pathologies exhibit typical time scales so small ($\approx 10^{-23}\, s$ for an electron) that were long considered harmless for all practical purposes. However the recent advances in ultra-intense laser technology \cite{di2012extremely, burton2014aspects} and related sophisticated numerical simulations have renewed interest in the field \cite{ji2014radiation, green2014transverse, vranic2016classical}.

The LAD equation is furthermore plagued by less severe problems usually called ambiguities. These are related to time-reversibility \cite{Rohrlich-1998, Zeh-1999, Rohrlich2001, Rovelli-2004}, a possible conflict with the Equivalence Principle \cite{parrott2002radiation, shariati1999equivalence}, infinite mass, etc.. In particular there is an energy balance paradox in the case of a uniformly accelerated charge for it does not seem possible to reach Larmor's formula from the LAD equation \cite{fulton1960classical, harpaz1998radiation, gupta1998radiation}.
The aim of this letter is twofold: on the one hand we reveal that the same discrepancy appears for arbitrary acceleration and on the other hand we show that the widely used Teitelboim and Schott interpretations \cite{Moylan1993, poisson1999introduction, eriksen2002electrodynamics, gal2002radiation, gal2004radiation, gralla2009rigorous, neto2015thinking}, related to the so-called \textit{bound field}, are ill defined and do not allow to fix the energy balance paradox even in the uniform acceleration case.

The amount of radiated energy per unit time of an accelerated charge is given by Larmor's formula,
\begin{equation}\label{Larmor}
\dot E_{rad} = - m\, \epsilon \, \ddot z^2,
\end{equation}
where $\epsilon = \frac{2 \, k\, e^2}{3mc^3}$ is a small time parameter and $e$ and $m$ being the electric charge and mass, respectively. The covariant four-acceleration reads $\ddot z^\mu$, where the charge coordinates $z^\mu(\tau)$ are functions of the proper time $\tau$ and the over dot stands for time derivative. 

To account for the lost energy, the charge is assumed to experience a damping reaction due to its own radiation. This results in the LAD equation of motion,
\begin{equation}\label{LAD}
m  \ddot z^\mu = F^\mu_{ext} + F^\mu_{LAD},
\end{equation}
where $F^\mu_{ext}$ is an exterior force orthogonal to $\dot z^\mu$, and
\begin{equation}\label{LAD-force}
F^\mu_{LAD} = m \epsilon (\dddot {z}^\mu + \frac{\ddot z^2}{c^2}\, \dot z^\mu),
\end{equation}
stands for the radiation damping force.
The first term is the Schott vector and the second is hereafter called the Larmor term. Note that unless otherwise indicated a vector means a Lorentz four-vector and that $c$ is kept for clarity when taking the non-relativistic limit.
\\

\paragraph*{Dirac's choice.}

The case of uniform (proper) acceleration is surprising. Instead of leading to simpler results, as one would expect, it yields more difficulties \cite{fulton1960classical, harpaz1998radiation}.
Uniform acceleration in the charge's rest frame is covariantly defined by 
\begin{equation}\label{uniform}
\dddot  z^\mu = - \frac{\ddot z^2}{c^2}\, \dot z^\mu,
\end{equation}
which implies 
$$ \dddot z.\ddot z=0 \implies \ddot z^2= {\vec a}_{proper}^2 = constant,$$ 
where ${\vec a}_{proper}$ is the uniform proper three-acceleration and so $\dot {\vec a}_{proper}=0$. Therefore the radiation force (\ref{LAD-force}) vanishes and the radiated energy (\ref{Larmor}) seems to come from nowhere\footnote{
This is so troublesome that important physicists like Pauli and Feynman have claimed there could be no radiation for uniform acceleration \cite{pauli1981theory, Feynman-gravity}. In addition, this might give rise to a conflict with the Equivalence Principle which locally equates uniform acceleration and a homogeneous gravitational field. A free charge on Earth would emit energy forever and this does not seem to happen. An intense work has been devoted to this problem, see \cite{higuchi1997static} and references therein. 
The accepted resolution to this conflict, due to Boulware, asserts that a uniformly accelerated charge does radiate, but such radiation cannot be detected by a comoving observer for falling outside her future cone \cite{boulware1980radiation}.}.
Hence there is no way to attain Larmor's formula from the LAD equation. This is well known and usually called the \textit{energy balance paradox}.
In what follows we evince that this difficulty holds in general, that is for arbitrary motion when the radiation force does not necessarily vanish.

In the standard treatment of classical electrodynamics, Larmor's formula (\ref{Larmor}) is derived from the Li\'enard-Wiechert potential.  The latter integrates the fields using the retarded Green function over the whole past history of the charge. In other words the Larmor's formula gives a globally defined four-momentum. On the same time one might ask how to reach the same result through the equation of motion, that is locally.

The usual way of getting an energy out of a four-force is to consider its zeroth component. However $F^0_{LAD}\, c=m\epsilon(\dddot z^0\, c + \ddot z^2)$ is clearly different from the radiated energy (\ref{Larmor}) because $\dddot z^0\neq 0$ for non vanishing three-acceleration.
Another way can be inferred from non-relativistic mechanics and consists in projecting the force on the velocity vector. However, due to the identity,
$$\dddot z.\dot z=-\ddot z^2 , $$
one has
$$m \epsilon (\dddot {z}^\mu + \frac{\ddot z^2}{c^2}\, \dot z^\mu) \ \dot z_\mu = 0.$$ 
Therefore the power of the radiation force is exactly vanishing when the radiated power is not (\ref{Larmor}).

To remedy this discrepancy, following Dirac, the contribution of the Schott term is usually eliminated from the radiated power expression. 
Dirac argued that for being a total derivative, the Schott term, $\dddot z^\mu$, is time-reversible and is therefore not propagating.

Notwithstanding, this argument is not valid because total derivation does not imply time-reversibility. The velocity $\dot z^\mu$ and acceleration $\ddot z^\mu$, for instance, are both total derivatives while the former is irreversible and the latter reversible.
In addition, for being an odd derivative of the position, the Schott term, $\dddot z^\mu$, is in fact time-irreversible, as is the velocity vector,  for example. 
Finally, writing the Larmor term under the form $\ddot z^2\dot z^\mu=-\dddot z.\dot z \dot z^\mu,$ allows to confer that the two terms of the radiation force (\ref{LAD-force}) are equally time-irreversible. That is, time-reversibility or not seems irrelevant in trying to attain Larmor's formula from the LAD equation.

Yet, even if we take on Dirac's argument for a moment, on his authority, we will face another problem in the non-relativistic limit ($c\to \infty$). Indeed, the Larmor term of the radiation force vanishes in this limit . Hence the radiation force reduces to the spatial component  of the Schott term and the LAD equation takes the form,
$$m\, \vec{a} =  \vec{f}  +   m \epsilon \,  \dot{\vec{a}}.$$  
Accordingly, Dirac's recipe does not make sense in the non-relativistic limit.
Note however that for the only cyclic configurations  (like rotations and oscillations which exhibit the identity $\dot{{a}}.{v}= \dot{\vec{a}}.\vec{v}= -\vec{a}^2$)
it is possible to recover the non-relativistic Larmor's formula but through the Schott term, $\dot E_{rad}= -m \epsilon \,  \dot{{a}} . {v} = m\, \epsilon\,  {a}^2$. 

Another illuminating example is the uniform acceleration motion in the laboratory frame\footnote{Note that this is different from uniform proper acceleration. Indeed the proper acceleration is related to the laboratory acceleration through $-\ddot z^2=\vec{a}_{proper}^2=\gamma^{6}\, (a.v^2+\gamma^{-2} \vec{a}^2)=\gamma^6\, \vec{a}^2$, the last identity holds for linear motion, note that here $\vec{a}=\vec{a}_{lab}$. So because of the Lorentz boost factor the proper and the laboratory accelerations cannot be both uniform at the same time.}.
This corresponds to set $\dot{\vec{a}}=0$ after expanding the LAD equation through $\dot z^\mu = \gamma(c,\vec{v})$, $\ddot z^\mu=\gamma^3\, \dot z^\mu + \gamma^2 (0, \vec{a})
=\gamma^4({a}.{v}/c, {a}.{v}/c^2 \, \vec{v} + \gamma^{-2}\, \vec{a})$ and so on, $\gamma$ being the Lorentz boost factor.
In this case the LAD equation (\ref{LAD}) reduces to\footnote{To our best knowledge this equation has never been written down eventhough it can be obtained straightforwardly from the covariant LAD equation (\ref{LAD}).}
\begin{equation}
m\, \gamma^3\,  \vec{a} =  \vec{f}  +  3 \epsilon \, m\, \gamma^6 \, \frac{a^2}{c^2}\,  \vec{v}.
\end{equation} 
The work done against the radiation force reads $3 \epsilon \, m\, \gamma^6 \, a^2 \, v^2/c^2$ which  is obviously different from the power given by Larmor's formula, $\dot E_{rad}=m\, \epsilon \gamma^6 {a}^2$.
In the non-relativistic limit the radiation force vanishes and one recovers the usual paradox of the uniform proper acceleration.

Consequently Dirac's choice does not seem to be well justified and one is left without knowing how to relate the radiated energy to the work done against the radiation reaction force.
\\

\paragraph*{Acceleration energy.}

The energy balance paradox appearing in the uniform acceleration case was revealed by Schott who was also the first to try to fix it  \cite{schott1915vi}. To do so he claimed that the term that now bears his name, $m\epsilon \, \dddot z^\mu$, in the radiation force (\ref{LAD-force}) should be considered as part of the charge and not of the propagating light. 
For this he promoted $m\epsilon \, \ddot z^\mu$ to what he called the acceleration momentum which was then included in the definition of the mechanical momentum of the charge
\begin{equation}\label{mech}
P_{mech}^\mu = m\, \dot z^\mu - m\epsilon \, \ddot z^\mu.
\end{equation}
The time variation of which is equated to the external force less  the radiated momentum, 
\begin{equation}\label{eq-mech}
\dot P_{mech}^\mu 
=  F^\mu_{ext} + m\epsilon \frac{\ddot z^2}{c^2}\dot z^\mu ,
\end{equation}
which is nothing but the LAD equation (\ref{LAD}). Hence, in Schott's model the energy radiated is partly (resp. fully) sourced by the acceleration energy of the charge for arbitrary (resp. uniform) acceleration.

The problem with Schott statement is double. First, the momentum defined by (\ref{mech}) is indefinite and is therefore not necessarily positive as it should. Indeed,  
$P_{mech}^2 = m^2\,(c^2 + \epsilon^2 \, \ddot z^2)
$ 
and $\ddot z^2<0$ for $\ddot z^\mu$ being spacelike. 
If one furthermore associates a mass to such momentum, $M^2=P_{mech}^2$, this mass is not constant but decreases with acceleration. Worse, for intense acceleration satisfying $\epsilon | \ddot z |>c$, the mass is imaginary and the charged particle is thus a tachyon \cite{kosyakov2007introduction}.

\begin{figure}[h]
\includegraphics[width = 8.7cm]{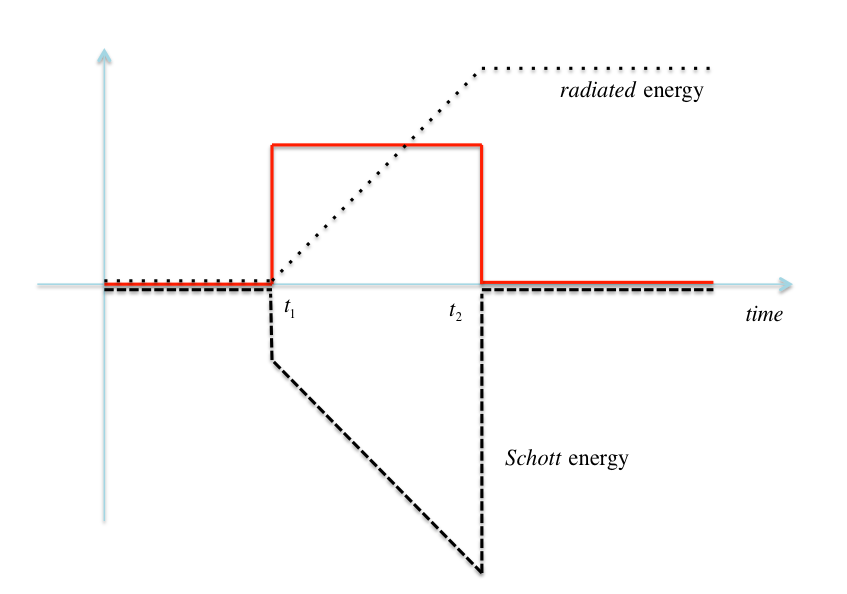}
  \caption{The continuous red line represents the uniform force/acceleration. The dotted line stands for the total energy radiated which reads from integrating Larmor's formula up to the instant $t$. 
Finally the dashed line represents  the Schott acceleration energy 
 which is supposed to be the source of the radiated energy. However, the jumps of the Schott energy at the instants $t_1$ and $t_2$ violate energy conservation.}
  \label{Fig1}
\end{figure}

Second, much more problematic, the acceleration momentum and thus Schott's mechanical momentum are not conserved. To see this let us consider a simple example with non relativistic motion (see Fig.\ref{Fig1}). Let a charge evolve in the laboratory inertial frame with constant velocity $v_1$ for times $t<t_1$ before being uniformly accelerated between $t_1$ and $t_2$. 
According to Larmor's interpretation, the mechanical energy of the charge goes from $E_{mech}(t<t_1) = m v_1^2/2$ to 
$
E_{mech}(t_1) =  m v_1^2/2 - m\epsilon \, a.v_1,
$
without any compensation for the decreased energy. One has a clear violation of energy conservation.
The acceleration energy ($m\epsilon \, a.v$) decreases linearly for $t_1<t<t_2$ and is supposed to account for the sum of the energy that is being radiated through, 
$$
m\epsilon \, a.v(t) = \int_{t_1}^{t} m\epsilon\, a^2 dt = m\epsilon\, a^2 \times (t-t_1).
$$
At the instant $t_2$ the mechanical energy attains
$
E_{mech}(t_2) = m v_2^2/2 - m\epsilon \, a.v_2,
$
when the external force is switched off. The next instant the acceleration term vanishes and thus the mechanical energy jumps to $E_{mech}(t>t_2) = m v_2^2/2 $ here again without a compensating source for the increased energy. This is the second violation of energy conservation.
Note finally that analysing the space component of Schott's mechanical momentum leads to the same jumps at $t_1$ and $t_2$ and thus to momentum conservation violation.

\paragraph*{Bound field.}
To avoid the radical change in the mechanical momentum definition the Schott term is more commonly considered as a third ingredient of the radiating charge system. The two others being the charge itself and the radiated wave. More explicitly, the Schott energy is thought as being related to the so-called near or bound field \cite{rohrlich1961}. Teitelboim was the first to give explicit relations \cite{teitelboim1970splitting}. He considered a charge interacting with its own electromagnetic field and split the Faraday tensor into velocity (I) and acceleration (II) parts
$ F^{\mu\nu} = F^{\mu\nu}_I + F^{\mu\nu}_{II}.$
Afterwards he computed the corresponding electromagnetic energy-momentum tensor which, for being quadratic in the fields, has three terms
$
T^{\mu\nu} = T^{\mu\nu}_{I,I} + T^{\mu\nu}_{I,II}+ T^{\mu\nu}_{II,II}.
$
The second term contains the interferences between the two parts of the Faraday field and since it does not propagate to infinity it is attached to the velocity part as\footnote{Note that attaching the interference term $T^{\mu\nu}_{I,II}$ with $T_{II,II}$ rather than $T_{I,I}$ leads to Dirac's original result 
$ \dot P^\mu_{I} = \frac{m\epsilon}{2\delta t} \ddot z^\mu ,$ and 
$ \dot P^\mu_{II} = - m\epsilon  ( \dddot z^\mu+ \frac{\ddot z^2}{c^2}\, \dot z^\mu),$
which nonetheless produces the same equation of motion.}
\begin{equation}\label{split}
T^{\mu\nu}_I = T^{\mu\nu}_{I,I} + T^{\mu\nu}_{I,II}, \qquad T^{\mu\nu}_{II}= T^{\mu\nu}_{II,II}.
\end{equation}
The corresponding momentum vectors are then computed using Dirac's method (which introduces an infinitesimal expansion time parameter $\delta t$), 
$$
\dot P^\mu_{I} = \frac{m\epsilon}{2\delta t} \ddot z^\mu -  m\epsilon \dddot z^\mu,
\qquad 
\dot P^\mu_{II} = - m\epsilon \frac{\ddot z^2}{c^2}\, \dot z^\mu.
$$
The first term of $\dot P^\mu_{I} $ is divergent in the limit $\delta t\to 0$, and is usually absorbed in the bare mass to obtain a renormalised and finite mass, $m=m_o + \frac{m\epsilon}{2\delta t}$.
Finally Teitelboim has defined what he called the four-momentum of the charge,
\begin{equation}\label{teit}
P^\mu = m_o \dot z^\mu + P^\mu_{I} = m\dot z^\mu - m\epsilon \ddot z^\mu.
\end{equation}
The equation of motion reads,
$
\dot P^\mu = F^\mu_{ext} - \dot P^\mu_{II},
$
and is the same as equation (\ref{eq-mech}) and thus (\ref{LAD}) as well.
Although the interpretation and terminology are different, the above expression (\ref{teit}) is exactly equal to Schott's mechanical momentum (\ref{mech}) and hence the same problems (indefinite and non conserved) hold for Teitelboim's momentum.

In addition, the non relativistic limit is problematic since the propagating term $\dot P_{II}^\mu$ vanishes and there is therefore no radiation at all.
Finally, Teitelboim's calculations are based on the asymptotic condition,
$$
\lim_{time\to\pm \infty} (motion) = uniform\ motion,
$$
which is restrictive and prevents considering the hyperbolic motion, for example. This obscures even more the bound field interpretation since it was precisely meant to fix the energy balance paradox emerging for uniform acceleration.

\paragraph*{Final remarks.}
This work reveals a new ambiguity in the LAD equation. Beyond known pathologies and ambiguities, we exhibit that there is a systematic energy balance discrepancy in the LAD equation. Moreover we show that the widely accepted treatment based on the bound field technique cannot fix this discrepancy even for uniform acceleration. The underlying reason is that the momentum defined by Schott and later by Teitelboim is not a legitimate four-momentum for being indefinite and non-conserved.

Note that using the Landau-Lifschitz equation in place of (\ref{LAD}) leads to the same ambiguity \cite{Landau-Lifshitz, baylis2002energy, griffiths2010abraham}. Known quantum models built to fix the radiating charge motion lead to the LAD equation in the classical limit and thus face the same problems \cite{Moniz, Higuchi-radiation}.
Moreover, beyond electromagnetic radiation, the same energy balance discrepancy  appears in gravity where the same technique of the near field is used to fix it \cite{poisson2004motion, gal2009radiation}.

Actually, Larmor has derived a formula for the radiated energy and momentum (as used in equation (\ref{eq-mech})) and not only for the energy, e.g. formula (\ref{Larmor}). The formula for the radiated four-momentum reads \cite{rohrlich1997dynamics},
\begin{equation}\label{Larmor-4}
\dot P_{rad}^\mu = - m\, \epsilon \, \ddot z^2\, \frac{\dot z^\mu}{c^2}.
\end{equation}
The radiated energy (\ref{Larmor}) corresponds to $\dot E_{rad}=c \, \dot P_{rad}^0$.
Considering the full radiated momentum it becomes more evident that the Larmor term in the radiation force should be the only source of the radiated momentum. The Schott term should hence be discarded when relating the radiation force and the radiation power. However, as we discussed, the arguments of Schott, Dirac, Teitelboim and others are not valid and do not allow to justify what seems astonishingly evident.

We would like to stress in what remains that extracting the dissipated energy from the equation of motion seems problematic for any possible Lorentz covariant dissipative equation of motion. Indeed, the dissipated (linear) momentum should be of the form
$$
\dot P^\mu_{diss} =\dot P_{diss}\, \frac{\dot z^\mu}{c^2},
$$
where $\dot P_{diss}=\dot P^\mu_{diss} \dot z^\mu$ is a scalar (thus Lorentz invariant) and stands for dissipated power (i.e. momentum's zero component).
Suppose furthermore that the reaction force due to the dissipation is given by $F_{diss}^\mu$ which enters the equation of motion as
$$
m \ddot z^\mu = F^\mu_{cons} + F_{diss}^\mu,
$$
where $F^\mu_{cons}$ is some external and conservative force.
For consistency $F_{diss}^\mu$ must be orthogonal to the velocity vector, i.e. $F_{diss}^\mu\, \dot z_\mu=0$. 
This is exactly what prevents us from using the usual nonrelativistic relation between force and energy. Hence the energy balance seems  generically paradoxical.
In fact, it appears that the dissipated momentum, $\dot P^\mu_{diss}$, and the damping force, $F^\mu_{diss}$, are orthogonal one another since the former is parallel while the latter is orthogonal to $\dot z^\mu$. Therefore the only way to relate them should rely on some englobing vector, say $H^\mu$, such that its parallel projection yields the dissipated momentum,
$$
H^\mu_{\parallel} = H_\nu\frac{\dot z^\nu \dot z^\mu}{c^2}
= \dot P_{diss}^\mu
$$
while the orthogonal projection yields the damping force, 
$$
H^\mu_{\perp} = H^\mu - H^\mu_{\parallel}  
=F^\mu_{diss}.
$$
Accordingly the dissipated momentum and the damping force are not directly equivalent but rather complementary, $H^\mu = F_{diss}^\mu+P_{diss}^\mu$.

The electromagnetic dissipation (radiation) case is obtained setting $H^\mu=m\epsilon \dddot z^\mu$. The parallel projection, $H^\mu_{\parallel}= -m\epsilon \ddot z^2\, \dot z^\mu/c^2$, provides the radiated power (zero component $\times$ c), $\dot P= H^0_{\parallel}\, c= -m\epsilon \ddot z^2$, which is exactly Larmor's formula (\ref{Larmor}). As to the orthogonal projection, $H^\mu_{\perp} = m\epsilon (\dddot z + \ddot z^2/c^2 \dot z^\mu)$, it yields the radiation damping force (\ref{LAD-force}).

Finally we would like to mention that these elements are being explored in the undergoing work \cite{sofiane-delay}. The latter is reconsidering the motion of radiating charges through a new and simple paradigm that might solve the present problem together with most known difficulties of the LAD equation.
\\

\paragraph*{Ackowledgment.} We thank Satheeshkumar V. H. for comments and the Brazilian MCTI (PCI program) and CNPq for financial support.

\baselineskip=10pt
\bibliography{Biblio}

\begin{thebibliography}{41}%
\makeatletter
\providecommand \@ifxundefined [1]{%
 \@ifx{#1\undefined}
}%
\providecommand \@ifnum [1]{%
 \ifnum #1\expandafter \@firstoftwo
 \else \expandafter \@secondoftwo
 \fi
}%
\providecommand \@ifx [1]{%
 \ifx #1\expandafter \@firstoftwo
 \else \expandafter \@secondoftwo
 \fi
}%
\providecommand \natexlab [1]{#1}%
\providecommand \enquote  [1]{``#1''}%
\providecommand \bibnamefont  [1]{#1}%
\providecommand \bibfnamefont [1]{#1}%
\providecommand \citenamefont [1]{#1}%
\providecommand \href@noop [0]{\@secondoftwo}%
\providecommand \href [0]{\begingroup \@sanitize@url \@href}%
\providecommand \@href[1]{\@@startlink{#1}\@@href}%
\providecommand \@@href[1]{\endgroup#1\@@endlink}%
\providecommand \@sanitize@url [0]{\catcode `\\12\catcode `\$12\catcode
  `\&12\catcode `\#12\catcode `\^12\catcode `\_12\catcode `\%12\relax}%
\providecommand \@@startlink[1]{}%
\providecommand \@@endlink[0]{}%
\providecommand \url  [0]{\begingroup\@sanitize@url \@url }%
\providecommand \@url [1]{\endgroup\@href {#1}{\urlprefix }}%
\providecommand \urlprefix  [0]{URL }%
\providecommand \Eprint [0]{\href }%
\providecommand \doibase [0]{http://dx.doi.org/}%
\providecommand \selectlanguage [0]{\@gobble}%
\providecommand \bibinfo  [0]{\@secondoftwo}%
\providecommand \bibfield  [0]{\@secondoftwo}%
\providecommand \translation [1]{[#1]}%
\providecommand \BibitemOpen [0]{}%
\providecommand \bibitemStop [0]{}%
\providecommand \bibitemNoStop [0]{.\EOS\space}%
\providecommand \EOS [0]{\spacefactor3000\relax}%
\providecommand \BibitemShut  [1]{\csname bibitem#1\endcsname}%
\let\auto@bib@innerbib\@empty
\bibitem [{\citenamefont {Rohrlick}(1965)}]{Rohrlick-book}%
  \BibitemOpen
  \bibfield  {author} {\bibinfo {author} {\bibfnamefont {F.}~\bibnamefont
  {Rohrlick}},\ }\href@noop {} {\emph {\bibinfo {title} {Classical Charged
  Particles}}},\ edited by\ \bibinfo {editor} {\bibfnamefont {M.}~\bibnamefont
  {Reading}}\ (\bibinfo  {publisher} {Addison-Wesley},\ \bibinfo {year}
  {1965})\BibitemShut {NoStop}%
\bibitem [{\citenamefont {Jackson}(1999)}]{jackson1999classical}%
  \BibitemOpen
  \bibfield  {author} {\bibinfo {author} {\bibfnamefont {J.~D.}\ \bibnamefont
  {Jackson}},\ }\href@noop {} {\emph {\bibinfo {title} {Classical
  electrodynamics}}}\ (\bibinfo  {publisher} {Wiley},\ \bibinfo {year}
  {1999})\BibitemShut {NoStop}%
\bibitem [{\citenamefont {Hammond}(2010)}]{Hammond-2010}%
  \BibitemOpen
  \bibfield  {author} {\bibinfo {author} {\bibfnamefont {R.}~\bibnamefont
  {Hammond}},\ }\href@noop {} {\bibfield  {journal} {\bibinfo  {journal}
  {EJTP}\ }\textbf {\bibinfo {volume} {7}},\ \bibinfo {pages} {221} (\bibinfo
  {year} {2010})}\BibitemShut {NoStop}%
\bibitem [{\citenamefont {Di~Piazza}\ \emph {et~al.}(2012)\citenamefont
  {Di~Piazza}, \citenamefont {M{\"u}ller}, \citenamefont {Hatsagortsyan},\ and\
  \citenamefont {Keitel}}]{di2012extremely}%
  \BibitemOpen
  \bibfield  {author} {\bibinfo {author} {\bibfnamefont {A.}~\bibnamefont
  {Di~Piazza}}, \bibinfo {author} {\bibfnamefont {C.}~\bibnamefont
  {M{\"u}ller}}, \bibinfo {author} {\bibfnamefont {K.}~\bibnamefont
  {Hatsagortsyan}}, \ and\ \bibinfo {author} {\bibfnamefont {C.}~\bibnamefont
  {Keitel}},\ }\href@noop {} {\bibfield  {journal} {\bibinfo  {journal}
  {Reviews of Modern Physics}\ }\textbf {\bibinfo {volume} {84}},\ \bibinfo
  {pages} {1177} (\bibinfo {year} {2012})}\BibitemShut {NoStop}%
\bibitem [{\citenamefont {Burton}\ and\ \citenamefont
  {Noble}(2014)}]{burton2014aspects}%
  \BibitemOpen
  \bibfield  {author} {\bibinfo {author} {\bibfnamefont {D.~A.}\ \bibnamefont
  {Burton}}\ and\ \bibinfo {author} {\bibfnamefont {A.}~\bibnamefont {Noble}},\
  }\href@noop {} {\bibfield  {journal} {\bibinfo  {journal} {Contemporary
  Physics}\ }\textbf {\bibinfo {volume} {55}},\ \bibinfo {pages} {110}
  (\bibinfo {year} {2014})}\BibitemShut {NoStop}%
\bibitem [{\citenamefont {Ji~LL}\ and\ \citenamefont
  {K}(2014)}]{ji2014radiation}%
  \BibitemOpen
  \bibfield  {author} {\bibinfo {author} {\bibfnamefont {K.~I. Y. S.~B.}\
  \bibnamefont {Ji~LL}, \bibfnamefont {Pukhov~A}}\ and\ \bibinfo {author}
  {\bibfnamefont {A.}~\bibnamefont {K}},\ }\href@noop {} {\bibfield  {journal}
  {\bibinfo  {journal} {Physical review letters}\ }\textbf {\bibinfo {volume}
  {112}},\ \bibinfo {pages} {145003} (\bibinfo {year} {2014})}\BibitemShut
  {NoStop}%
\bibitem [{\citenamefont {Green}\ and\ \citenamefont
  {Harvey}(2014)}]{green2014transverse}%
  \BibitemOpen
  \bibfield  {author} {\bibinfo {author} {\bibfnamefont {D.}~\bibnamefont
  {Green}}\ and\ \bibinfo {author} {\bibfnamefont {C.}~\bibnamefont {Harvey}},\
  }\href@noop {} {\bibfield  {journal} {\bibinfo  {journal} {Physical Review
  Letters}\ }\textbf {\bibinfo {volume} {112}},\ \bibinfo {pages} {164801}
  (\bibinfo {year} {2014})}\BibitemShut {NoStop}%
\bibitem [{\citenamefont {Vranic}\ \emph {et~al.}(2016)\citenamefont {Vranic},
  \citenamefont {Martins}, \citenamefont {Fonseca},\ and\ \citenamefont
  {Silva}}]{vranic2016classical}%
  \BibitemOpen
  \bibfield  {author} {\bibinfo {author} {\bibfnamefont {M.}~\bibnamefont
  {Vranic}}, \bibinfo {author} {\bibfnamefont {J.~L.}\ \bibnamefont {Martins}},
  \bibinfo {author} {\bibfnamefont {R.~A.}\ \bibnamefont {Fonseca}}, \ and\
  \bibinfo {author} {\bibfnamefont {L.~O.}\ \bibnamefont {Silva}},\ }\href@noop
  {} {\bibfield  {journal} {\bibinfo  {journal} {Computer Physics
  Communications}\ }\textbf {\bibinfo {volume} {204}},\ \bibinfo {pages} {141}
  (\bibinfo {year} {2016})}\BibitemShut {NoStop}%
\bibitem [{\citenamefont {Rohrlich}(1998)}]{Rohrlich-1998}%
  \BibitemOpen
  \bibfield  {author} {\bibinfo {author} {\bibfnamefont {F.}~\bibnamefont
  {Rohrlich}},\ }\href@noop {} {\bibfield  {journal} {\bibinfo  {journal}
  {Foundations of Physics,}\ }\textbf {\bibinfo {volume} {28}} (\bibinfo {year}
  {1998})}\BibitemShut {NoStop}%
\bibitem [{\citenamefont {Zeh}(1999)}]{Zeh-1999}%
  \BibitemOpen
  \bibfield  {author} {\bibinfo {author} {\bibfnamefont {H.~D.}\ \bibnamefont
  {Zeh}},\ }\href@noop {} {\bibfield  {journal} {\bibinfo  {journal}
  {Foundations of Physics Letters,}\ }\textbf {\bibinfo {volume} {12}}
  (\bibinfo {year} {1999})}\BibitemShut {NoStop}%
\bibitem [{\citenamefont {Rohrlich}(2000)}]{Rohrlich2001}%
  \BibitemOpen
  \bibfield  {author} {\bibinfo {author} {\bibfnamefont {F.}~\bibnamefont
  {Rohrlich}},\ }\href {\doibase
  http://dx.doi.org/10.1016/S1355-2198(99)00030-1} {\bibfield  {journal}
  {\bibinfo  {journal} {Studies in Hist. and Phil. of Mod. Phys.}\ }\textbf
  {\bibinfo {volume} {31}},\ \bibinfo {pages} {1 } (\bibinfo {year}
  {2000})}\BibitemShut {NoStop}%
\bibitem [{\citenamefont {Rovelli}(2004)}]{Rovelli-2004}%
  \BibitemOpen
  \bibfield  {author} {\bibinfo {author} {\bibfnamefont {C.}~\bibnamefont
  {Rovelli}},\ }\href@noop {} {\bibfield  {journal} {\bibinfo  {journal}
  {Studies in History and Philosophy of Science}\ }\textbf {\bibinfo {volume}
  {35}},\ \bibinfo {pages} {397} (\bibinfo {year} {2004})}\BibitemShut
  {NoStop}%
\bibitem [{\citenamefont {Parrott}(2002)}]{parrott2002radiation}%
  \BibitemOpen
  \bibfield  {author} {\bibinfo {author} {\bibfnamefont {S.}~\bibnamefont
  {Parrott}},\ }\href@noop {} {\bibfield  {journal} {\bibinfo  {journal}
  {Foundations of Physics}\ }\textbf {\bibinfo {volume} {32}},\ \bibinfo
  {pages} {407} (\bibinfo {year} {2002})}\BibitemShut {NoStop}%
\bibitem [{\citenamefont {Shariati}\ and\ \citenamefont
  {Khorrami}(1999)}]{shariati1999equivalence}%
  \BibitemOpen
  \bibfield  {author} {\bibinfo {author} {\bibfnamefont {A.}~\bibnamefont
  {Shariati}}\ and\ \bibinfo {author} {\bibfnamefont {M.}~\bibnamefont
  {Khorrami}},\ }\href@noop {} {\bibfield  {journal} {\bibinfo  {journal}
  {Foundations of Physics Letters}\ }\textbf {\bibinfo {volume} {12}},\
  \bibinfo {pages} {427} (\bibinfo {year} {1999})}\BibitemShut {NoStop}%
\bibitem [{\citenamefont {Fulton}\ and\ \citenamefont
  {Rohrlich}(1960)}]{fulton1960classical}%
  \BibitemOpen
  \bibfield  {author} {\bibinfo {author} {\bibfnamefont {T.}~\bibnamefont
  {Fulton}}\ and\ \bibinfo {author} {\bibfnamefont {F.}~\bibnamefont
  {Rohrlich}},\ }\href@noop {} {\bibfield  {journal} {\bibinfo  {journal}
  {Annals of Physics}\ }\textbf {\bibinfo {volume} {9}},\ \bibinfo {pages}
  {499} (\bibinfo {year} {1960})}\BibitemShut {NoStop}%
\bibitem [{\citenamefont {Harpaz}\ and\ \citenamefont
  {Soker}(1998)}]{harpaz1998radiation}%
  \BibitemOpen
  \bibfield  {author} {\bibinfo {author} {\bibfnamefont {A.}~\bibnamefont
  {Harpaz}}\ and\ \bibinfo {author} {\bibfnamefont {N.}~\bibnamefont {Soker}},\
  }\href@noop {} {\bibfield  {journal} {\bibinfo  {journal} {GR and Grav.}\
  }\textbf {\bibinfo {volume} {30}},\ \bibinfo {pages} {1217} (\bibinfo {year}
  {1998})}\BibitemShut {NoStop}%
\bibitem [{\citenamefont {Gupta}\ and\ \citenamefont
  {Padmanabhan}(1998)}]{gupta1998radiation}%
  \BibitemOpen
  \bibfield  {author} {\bibinfo {author} {\bibfnamefont {A.}~\bibnamefont
  {Gupta}}\ and\ \bibinfo {author} {\bibfnamefont {T.}~\bibnamefont
  {Padmanabhan}},\ }\href@noop {} {\bibfield  {journal} {\bibinfo  {journal}
  {Physical Review D}\ }\textbf {\bibinfo {volume} {57}},\ \bibinfo {pages}
  {7241} (\bibinfo {year} {1998})}\BibitemShut {NoStop}%
\bibitem [{\citenamefont {Moylan}(1993)}]{Moylan1993}%
  \BibitemOpen
  \bibfield  {author} {\bibinfo {author} {\bibfnamefont {P.}~\bibnamefont
  {Moylan}},\ }\href {\doibase 10.1007/BF00675015} {\bibfield  {journal}
  {\bibinfo  {journal} {International Journal of Theoretical Physics}\ }\textbf
  {\bibinfo {volume} {32}},\ \bibinfo {pages} {2031} (\bibinfo {year}
  {1993})}\BibitemShut {NoStop}%
\bibitem [{\citenamefont {Poisson}(1999)}]{poisson1999introduction}%
  \BibitemOpen
  \bibfield  {author} {\bibinfo {author} {\bibfnamefont {E.}~\bibnamefont
  {Poisson}},\ }\href@noop {} {\bibfield  {journal} {\bibinfo  {journal} {arXiv
  preprint gr-qc/9912045}\ } (\bibinfo {year} {1999})}\BibitemShut {NoStop}%
\bibitem [{\citenamefont {Eriksen}\ and\ \citenamefont
  {Gr{\o}n}(2002)}]{eriksen2002electrodynamics}%
  \BibitemOpen
  \bibfield  {author} {\bibinfo {author} {\bibfnamefont {E.}~\bibnamefont
  {Eriksen}}\ and\ \bibinfo {author} {\bibfnamefont {{\O}.}~\bibnamefont
  {Gr{\o}n}},\ }\href@noop {} {\bibfield  {journal} {\bibinfo  {journal}
  {Annals of Physics}\ }\textbf {\bibinfo {volume} {297}},\ \bibinfo {pages}
  {243} (\bibinfo {year} {2002})}\BibitemShut {NoStop}%
\bibitem [{\citenamefont {Gal'Tsov}(2002)}]{gal2002radiation}%
  \BibitemOpen
  \bibfield  {author} {\bibinfo {author} {\bibfnamefont {D.~V.}\ \bibnamefont
  {Gal'Tsov}},\ }\href@noop {} {\bibfield  {journal} {\bibinfo  {journal}
  {Physical Review D}\ }\textbf {\bibinfo {volume} {66}},\ \bibinfo {pages}
  {025016} (\bibinfo {year} {2002})}\BibitemShut {NoStop}%
\bibitem [{\citenamefont {Gal'tsov}\ and\ \citenamefont
  {Spirin}(2004)}]{gal2004radiation}%
  \BibitemOpen
  \bibfield  {author} {\bibinfo {author} {\bibfnamefont {D.~V.}\ \bibnamefont
  {Gal'tsov}}\ and\ \bibinfo {author} {\bibfnamefont {P.}~\bibnamefont
  {Spirin}},\ }\href@noop {} {\bibfield  {journal} {\bibinfo  {journal} {arXiv
  preprint hep-th/0405121}\ } (\bibinfo {year} {2004})}\BibitemShut {NoStop}%
\bibitem [{\citenamefont {Gralla}\ \emph {et~al.}(2009)\citenamefont {Gralla},
  \citenamefont {Harte},\ and\ \citenamefont {Wald}}]{gralla2009rigorous}%
  \BibitemOpen
  \bibfield  {author} {\bibinfo {author} {\bibfnamefont {S.~E.}\ \bibnamefont
  {Gralla}}, \bibinfo {author} {\bibfnamefont {A.~I.}\ \bibnamefont {Harte}}, \
  and\ \bibinfo {author} {\bibfnamefont {R.~M.}\ \bibnamefont {Wald}},\
  }\href@noop {} {\bibfield  {journal} {\bibinfo  {journal} {Physical Review
  D}\ }\textbf {\bibinfo {volume} {80}},\ \bibinfo {pages} {024031} (\bibinfo
  {year} {2009})}\BibitemShut {NoStop}%
\bibitem [{\citenamefont {Neto}\ and\ \citenamefont
  {Helayel-Neto}(2015)}]{neto2015thinking}%
  \BibitemOpen
  \bibfield  {author} {\bibinfo {author} {\bibfnamefont {J.}~\bibnamefont
  {Neto}}\ and\ \bibinfo {author} {\bibfnamefont {J.~A.}\ \bibnamefont
  {Helayel-Neto}},\ }\href@noop {} {\bibfield  {journal} {\bibinfo  {journal}
  {arXiv:1504.04669}\ } (\bibinfo {year} {2015})}\BibitemShut {NoStop}%
\bibitem [{\citenamefont {Pauli}(1981)}]{pauli1981theory}%
  \BibitemOpen
  \bibfield  {author} {\bibinfo {author} {\bibfnamefont {W.}~\bibnamefont
  {Pauli}},\ }\href@noop {} {\emph {\bibinfo {title} {Theory of relativity}}}\
  (\bibinfo  {publisher} {Courier Corporation},\ \bibinfo {year}
  {1981})\BibitemShut {NoStop}%
\bibitem [{\citenamefont {Feynman}(1995)}]{Feynman-gravity}%
  \BibitemOpen
  \bibfield  {author} {\bibinfo {author} {\bibfnamefont {R.}~\bibnamefont
  {Feynman}},\ }\href@noop {} {\emph {\bibinfo {title} {Feynman Lectures On
  Gravitation}}},\ edited by\ \bibinfo {editor} {\bibfnamefont
  {F.}~\bibnamefont {in~Physics~S}}\ (\bibinfo  {publisher} {Frontiers in
  Physics S},\ \bibinfo {year} {1995})\BibitemShut {NoStop}%
\bibitem [{\citenamefont {Higuchi}\ \emph {et~al.}(1997)\citenamefont
  {Higuchi}, \citenamefont {Matsas},\ and\ \citenamefont
  {Sudarsky}}]{higuchi1997static}%
  \BibitemOpen
  \bibfield  {author} {\bibinfo {author} {\bibfnamefont {A.}~\bibnamefont
  {Higuchi}}, \bibinfo {author} {\bibfnamefont {G.~E.}\ \bibnamefont {Matsas}},
  \ and\ \bibinfo {author} {\bibfnamefont {D.}~\bibnamefont {Sudarsky}},\
  }\href@noop {} {\bibfield  {journal} {\bibinfo  {journal} {Physical Review
  D}\ }\textbf {\bibinfo {volume} {56}},\ \bibinfo {pages} {R6071} (\bibinfo
  {year} {1997})}\BibitemShut {NoStop}%
\bibitem [{\citenamefont {Boulware}(1980)}]{boulware1980radiation}%
  \BibitemOpen
  \bibfield  {author} {\bibinfo {author} {\bibfnamefont {D.~G.}\ \bibnamefont
  {Boulware}},\ }\href@noop {} {\bibfield  {journal} {\bibinfo  {journal}
  {Annals of Physics}\ }\textbf {\bibinfo {volume} {124}},\ \bibinfo {pages}
  {169} (\bibinfo {year} {1980})}\BibitemShut {NoStop}%
\bibitem [{\citenamefont {Schott}(1915)}]{schott1915vi}%
  \BibitemOpen
  \bibfield  {author} {\bibinfo {author} {\bibfnamefont {G.}~\bibnamefont
  {Schott}},\ }\href@noop {} {\bibfield  {journal} {\bibinfo  {journal} {The
  London, Edinburgh, and Dublin Philosophical Magazine and Journal of Science}\
  }\textbf {\bibinfo {volume} {29}},\ \bibinfo {pages} {49} (\bibinfo {year}
  {1915})}\BibitemShut {NoStop}%
\bibitem [{\citenamefont {Kosyakov}(2007)}]{kosyakov2007introduction}%
  \BibitemOpen
  \bibfield  {author} {\bibinfo {author} {\bibfnamefont {B.}~\bibnamefont
  {Kosyakov}},\ }\href@noop {} {\emph {\bibinfo {title} {Introduction to the
  classical theory of particles and fields}}}\ (\bibinfo  {publisher} {Springer
  Science},\ \bibinfo {year} {2007})\BibitemShut {NoStop}%
\bibitem [{\citenamefont {Rohrlich}(1961)}]{rohrlich1961}%
  \BibitemOpen
  \bibfield  {author} {\bibinfo {author} {\bibfnamefont {F.}~\bibnamefont
  {Rohrlich}},\ }\href@noop {} {\bibfield  {journal} {\bibinfo  {journal}
  {Annals of Physics}\ }\textbf {\bibinfo {volume} {13}},\ \bibinfo {pages}
  {93} (\bibinfo {year} {1961})}\BibitemShut {NoStop}%
\bibitem [{\citenamefont {Teitelboim}(1970)}]{teitelboim1970splitting}%
  \BibitemOpen
  \bibfield  {author} {\bibinfo {author} {\bibfnamefont {C.}~\bibnamefont
  {Teitelboim}},\ }\href@noop {} {\bibfield  {journal} {\bibinfo  {journal}
  {Physical Review D}\ }\textbf {\bibinfo {volume} {1}},\ \bibinfo {pages}
  {1572} (\bibinfo {year} {1970})}\BibitemShut {NoStop}%
\bibitem [{\citenamefont {Landau}\ and\ \citenamefont
  {Lifshitz}(1975)}]{Landau-Lifshitz}%
  \BibitemOpen
  \bibfield  {author} {\bibinfo {author} {\bibfnamefont {L.}~\bibnamefont
  {Landau}}\ and\ \bibinfo {author} {\bibfnamefont {E.}~\bibnamefont
  {Lifshitz}},\ }\href@noop {} {\emph {\bibinfo {title} {The Classical Theory
  of Fields}}},\ Vol.\ \bibinfo {volume} {4th Ed.}\ (\bibinfo  {publisher}
  {Butterworth-Heinemann},\ \bibinfo {year} {1975})\BibitemShut {NoStop}%
\bibitem [{\citenamefont {Baylis}\ and\ \citenamefont
  {Huschilt}(2002)}]{baylis2002energy}%
  \BibitemOpen
  \bibfield  {author} {\bibinfo {author} {\bibfnamefont {W.}~\bibnamefont
  {Baylis}}\ and\ \bibinfo {author} {\bibfnamefont {J.}~\bibnamefont
  {Huschilt}},\ }\href@noop {} {\bibfield  {journal} {\bibinfo  {journal}
  {Physics Letters A}\ }\textbf {\bibinfo {volume} {301}},\ \bibinfo {pages}
  {7} (\bibinfo {year} {2002})}\BibitemShut {NoStop}%
\bibitem [{\citenamefont {Griffiths}\ \emph {et~al.}(2010)\citenamefont
  {Griffiths}, \citenamefont {Proctor},\ and\ \citenamefont
  {Schroeter}}]{griffiths2010abraham}%
  \BibitemOpen
  \bibfield  {author} {\bibinfo {author} {\bibfnamefont {D.~J.}\ \bibnamefont
  {Griffiths}}, \bibinfo {author} {\bibfnamefont {T.~C.}\ \bibnamefont
  {Proctor}}, \ and\ \bibinfo {author} {\bibfnamefont {D.~F.}\ \bibnamefont
  {Schroeter}},\ }\href@noop {} {\bibfield  {journal} {\bibinfo  {journal}
  {American Journal of Physics}\ }\textbf {\bibinfo {volume} {78}},\ \bibinfo
  {pages} {391} (\bibinfo {year} {2010})}\BibitemShut {NoStop}%
\bibitem [{\citenamefont {Moniz}\ and\ \citenamefont {Sharp}(1977)}]{Moniz}%
  \BibitemOpen
  \bibfield  {author} {\bibinfo {author} {\bibfnamefont {E.~J.}\ \bibnamefont
  {Moniz}}\ and\ \bibinfo {author} {\bibfnamefont {D.~H.}\ \bibnamefont
  {Sharp}},\ }\href@noop {} {\bibfield  {journal} {\bibinfo  {journal} {Phys.
  Rev. D}\ }\textbf {\bibinfo {volume} {15}} (\bibinfo {year}
  {1977})}\BibitemShut {NoStop}%
\bibitem [{\citenamefont {Higuchi}\ and\ \citenamefont
  {Martin}(2004)}]{Higuchi-radiation}%
  \BibitemOpen
  \bibfield  {author} {\bibinfo {author} {\bibfnamefont {A.}~\bibnamefont
  {Higuchi}}\ and\ \bibinfo {author} {\bibfnamefont {G.~D.~R.}\ \bibnamefont
  {Martin}},\ }\href@noop {} {\bibfield  {journal} {\bibinfo  {journal} {Phys.
  Rev. D}\ }\textbf {\bibinfo {volume} {70}} (\bibinfo {year}
  {2004})}\BibitemShut {NoStop}%
\bibitem [{\citenamefont {Poisson}(2004)}]{poisson2004motion}%
  \BibitemOpen
  \bibfield  {author} {\bibinfo {author} {\bibfnamefont {E.}~\bibnamefont
  {Poisson}},\ }\href@noop {} {\bibfield  {journal} {\bibinfo  {journal}
  {Living Rev. Rel}\ }\textbf {\bibinfo {volume} {7}},\ \bibinfo {pages} {7}
  (\bibinfo {year} {2004})}\BibitemShut {NoStop}%
\bibitem [{\citenamefont {Gal'tsov}(2009)}]{gal2009radiation}%
  \BibitemOpen
  \bibfield  {author} {\bibinfo {author} {\bibfnamefont {D.}~\bibnamefont
  {Gal'tsov}},\ }in\ \href@noop {} {\emph {\bibinfo {booktitle} {Mass and
  Motion in General Relativity}}}\ (\bibinfo  {publisher} {Springer},\ \bibinfo
  {year} {2009})\ pp.\ \bibinfo {pages} {367--393}\BibitemShut {NoStop}%
\bibitem [{\citenamefont {Rohrlich}(1997)}]{rohrlich1997dynamics}%
  \BibitemOpen
  \bibfield  {author} {\bibinfo {author} {\bibfnamefont {F.}~\bibnamefont
  {Rohrlich}},\ }\href@noop {} {\bibfield  {journal} {\bibinfo  {journal}
  {American Journal of Physics}\ }\textbf {\bibinfo {volume} {65}},\ \bibinfo
  {pages} {1051} (\bibinfo {year} {1997})}\BibitemShut {NoStop}%
\bibitem [{\citenamefont {Faci}\ and\ \citenamefont
  {Novello}(2016)}]{sofiane-delay}%
  \BibitemOpen
  \bibfield  {author} {\bibinfo {author} {\bibfnamefont {S.}~\bibnamefont
  {Faci}}\ and\ \bibinfo {author} {\bibfnamefont {M.}~\bibnamefont {Novello}},\
  }\href@noop {} {\bibfield  {journal} {\bibinfo  {journal} {An infinitesimal
  time-delay mechanism to describe the motion of a radiating charge}\ }
  (\bibinfo {year} {2016})}\BibitemShut {NoStop}%
\end{thebibliography}%

\end{document}